\documentclass{article}
\usepackage{graphicx} % Required for inserting images
\usepackage{natbib}
\usepackage{subcaption}
\usepackage{setspace}
\usepackage[colorlinks=true]{hyperref}

\title{Multi-Study Causal Forest (MCF): A flexible framework for data borrowing in the presence of varying treatment effect heterogeneity} 

\author{Ashwini Venkatasubramaniam$^\mathsection$ and Julian Wolfson$^\dagger$\\[4pt]
}
\date{%
   \textit{$^\mathsection$GlaxoSmithKline, Stevenage, UK}\\%
    \textit{$^\dagger$Division of Biostatistics \& Health Data Science, School of Public Health, University of Minnesota, Minneapolis, USA}\\%
    \today
}

\begin{document}
\maketitle

\begin{abstract}
Tailoring treatment assignment to specific individuals can improve the health outcomes, but a single study may offer inadequate information for this purpose. The ability to leverage information from an auxiliary data source deemed to be `most similar' to a primary data source has been shown to improve estimates of treatment effects. In this paper, we introduce a framework, the Multi-Study Causal Forest (MCF), to borrow individual patient-level data from an auxiliary data source in the presence of `varying sources' of treatment effect heterogeneity. We utilise a simulation study to demonstrate the superiority of the MCF in the presence of varying treatment allocation models (between-study heterogeneity) in addition to being able to account for the presence of within-study heterogeneity. This approach can combine data from randomised controlled trials, observational studies or a combination of both. We illustrate using Breast cancer data that the MCF performs favourably compared to an existing methodology in the presence of varying sources of (both between and within) heterogeneity. 

\end{abstract}

\section{Introduction}

Treatment effects often vary across patients based on unique individual characteristics, and quantifying this heterogeneity is integral to the development of personalised treatment strategies. The key causal estimand of interest in this context is the conditional average treatment effect (CATE)—the average patient’s response to a given treatment conditional on their characteristics. However, many studies (both randomised and observational) are either too small or too noisy to precisely estimate CATEs. A promising approach to improve the estimation of heterogeneous treatment effects is to leverage data from other studies that apply the same (or similar) treatment to a related population. The principle behind such data integration aligns closely with the recent literature on transfer learning for causal inference and domain adaptation, where the goal is to borrow strength from auxiliary datasets while acknowledging that distributions, patient populations, or study designs may differ between sources. This perspective connects with frameworks like transportability \citep{bareinboim2014transportability}, invariant causal prediction \citep{peters2016causal}, and balanced representation learning \citep{johansson2016learning}, all of which seek to understand when and how causal information can be reliably "transferred" across different data environments.

From a Bayesian perspective, various methods have been proposed to incorporate supplemental data into the analysis of a primary data source. Traditional borrowing approaches—both static \citep{viele2014use} and dynamic \citep{han2017covariate, chen2011bayesian}—evaluate and incorporate external information primarily at the level of marginal or average treatment effects. These methods, which may treat data sources as exchangeable (or nearly so), connect to the idea of adaptive pooling in the transfer learning literature, where the "amount" of information borrowed is tuned based on how well the external source aligns with the primary target domain. Dynamic borrowing methods have expanded to handle multiple data sources with varying degrees of similarity. For instance, Multisource Exchangeability Models (MEMs) \citep{kaizer2018bayesian}, utilizing Bayesian Model Averaging \citep{hoeting1999bayesian}, can integrate data from a mixture of exchangeable and non-exchangeable sources. However, these approaches typically focus on borrowing strength for the average treatment effect (ATE), implicitly assuming that treatment effect heterogeneity is limited or that the same pattern of effect modification holds across data sources.

Recent advances in transfer learning for causal inference explicitly address these issues by focusing on CATEs rather than ATEs. Approaches such as those proposed by \cite{kotalik2021dynamic} extend MEM frameworks to accommodate heterogeneity, though their primary focus remains on the ATE in the primary population. Complementing these efforts, a rich array of non-parametric methods developed for single-study scenarios—such as Meta-Learners \citep{kunzel2019metalearners}, the R-learner \citep{nie2021quasi}, DR (Doubly Robust)-Learner \citep{kennedy2023towards}, Causal Forest \citep{wager2018estimation, powers2018some}, Bayesian Causal Forest (BCF) \citep{hahn2020bayesian}, and Model-based Forests \citep{dandl2024makes}—enable flexible modeling of treatment effect heterogeneity. More recently, researchers have begun to consider how non-parametric frameworks can be extended to multi-study or multi-domain settings, thus embracing the essence of transfer learning in the context of heterogeneous effect estimation. For example, tree-based ensemble frameworks for federated learning settings \citep{tan2022tree} and aggregate Bayesian Causal Forests \citep{thal2024aggregate} illustrate how multi-study integration can be achieved even with limited data sharing. In scenarios where patient-level data are available across multiple trials, generalizations of the single-study methods have been proposed \citep{brantner2024comparison}, and comprehensive reviews of such methods highlight the various strategies for combining information across heterogeneous domains \citep{brantner2023methods}. Relatedly, the multi-study R-Learner \citep{shyr2023multi} harnesses data-adaptive objective functions to borrow information from diverse data sources (RCTs and/or observational studies).

In line with this evolving literature, we introduce a framework—the Multi-Study Causal Forest (MCF)—to borrow individual patient-level data from an auxiliary data source with the goal of improving the estimation of heterogeneous causal treatment effects in a primary data source. We do so by: (1) evaluating the similarity of treatment effect heterogeneity patterns between these sources and the primary study population, and (2) determining how much information to pool from each supplementary source based on that similarity. This approach seeks to minimize assumptions and only requires that exchangeability of data sources holds once we condition on observed covariates, aligning naturally with the "shared structure" premise underlying many transfer learning techniques. By using a Causal Forest framework, we allow for non-parametric relationships that can capture complex patterns of treatment effect heterogeneity across multiple data domains. Through simulation and an illustrative application to openly available breast cancer datasets (combining an RCT as the primary source with a supplementary observational dataset), we demonstrate how our approach, which is both computationally efficient and straightforward to implement in standard statistical software, can flexibly integrate and adaptively borrow strength to more precisely estimate heterogeneous treatment effects. 

\section{Method}

\subsection{Setup}

For simplicity, consider a scenario where we have $n_p$ individuals in a `primary' study $s_{pr}$ and $n_a$ individuals from a single `auxiliary' study $s_a$ with a commonly defined outcome $Y$ and binary treatment $Z$. Let the training and test datasets derived from the primary dataset be denoted by $s_{tr}$ and $s_{te}$. A simple combination of the training and auxiliary dataset may be denoted as $s_{tr+a}$. For each $i \in 1, \dots, n_p, \dots, n_{p} + n_a$, let $Y_i$ denote the outcome, $Z_i \in \{0, 1\}$ be the treatment indicator and $S_i \in \{0, 1 \}$  denote whether the individual is from the primary (=0) or auxiliary (=1) study. Using the potential outcomes framework \citep{rubin1974estimating}, let $\{Y_i(1), Y_i(0)\}$ represent the potential outcomes that would have been observed given treatment allocation, $Z_i = 1$ (for treatment) and $Z_i = 0$ (for control), and $Y_i(1) - Y_i(0)$ represent the (causal) treatment effect of $Z$. Let $\mathbf{X}_i$ represent the matrix of baseline covariates. The propensity score is defined as $e_i(x) = Pr(Z_i=1| \mathbf{X}_i = x)$, to determine the probability of receiving a treatment assignment given the baseline covariates.

In many contexts, a key goal is to estimate the conditional average treatment effect (CATE), defined by $\tau(x) = E[Y_i(1) - Y_i(0) | X_i = x]$, a quantity which can be used to characterize (causal) treatment effect heterogeneity. Here, we are specifically interested in estimating the CATE within the primary study population,  $\tau_{p}(x) = E[Y_i(1) - Y_i(0) | X_i = x, S_i = 0]$. In the presence of auxiliary data, it is reasonable to ask whether those data can be leveraged to improve precision in estimation of $\tau_p$. Intuitively, we should be able to borrow information from the auxiliary source if $\tau_p = \tau_a \equiv E[Y_i(1) - Y_i(0) | X_i = x, S_i = 1]$, i.e., the CATEs in the primary and auxiliary data source are exchangeable. 

Suppose that the data in the primary data set derive from the joint distribution $F_p(y,z,x)$ and the auxiliary data derive from $F_a(y,z,x)$. In an idealized setting where the data in both studies derive from the same underlying joint distribution $F(y,z,x)$, the treatment and outcome were identically defined, implemented, and measured in both the primary and auxiliary data, and the covariate matrix $\mathbf{X}$ contained all the covariates driving heterogeneity of the effect of $Z$ on $Y$, then $\tau_p = \tau_a$ by definition. In practice, however, there are several reasons why we may have $\tau_p \neq \tau_a$. The first reason for non-exchangeability is modest differences in how the treatment is delivered. For example, the primary and auxiliary studies may involve the same medication delivered at different dosages or via different mechanisms (e.g., pill vs. injection).  It may be reasonable to believe that the drivers of effect heterogeneity are unchanged when minor modifications are made to how the treatment is delivered, but this is not guaranteed. The second reason for non-exchangeability is differences in how the outcome is defined and measured. As with treatment, when outcomes are collected in a similar but not identical manner across the primary and auxiliary studies, there is a possibility that CATE exchangeability will not hold. A final reason for potential non-exchangeability of CATEs is if key covariates which drive effect heterogeneity are unmeasured. In this case, while the fully conditional CATEs which account for all heterogeneity-driving covariates remain exchangeable, the partially conditional CATEs which include only measured covariates may not match as they marginalize over different unmeasured covariate distributions.

% Within a given study, the estimand of interest is the conditional average treatment effect (CATE) and this study specific CATE is denoted by $\tau_s(x)$, where $\tau_s(x) = E[Y_i(1) - Y_i(0) | X_i = x, S_i = s]$. Accordingly, for the primary dataset $\tau_{s_p}(x) = E[Y_i(1) - Y_i(0) | X_i = x, S_i = s_p]$ and for the auxiliary dataset $\tau_{s_a}(x) = E[Y_i(1) - Y_i(0) | X_i = x, S_i = s_a]$ However, we are interested in determining how best to pool datasets when multiple datasets are available and so seek to compute the overall CATE (not specific to a given study $s$). Here $\tau(x) = E[Y_i(1) - Y_i(0) | X_i = x]$. 

%Covariate-adjusted exchangeability is defined between the primary source $s_p$ and an auxiliary data $s_a$ as $\tau_{s_p}(x) = \tau_{s_a}(x)$. Therefore, using a given set of covariates $\mathbf{X}_p$ from the primary source, $\tau_{s_p}(\mathbf{X}_p) = \tau_{s_a}(\mathbf{X}_p)$. In other words, adjusting for the covariate distributions between the primary study and the auxiliary study would account for any differences between the equivalent functions $\tau_{s_p}(x)$ and $\tau_{s_a}(x)$. However, in the presence of treatment effect heterogeneity, it is unlikely to have two data sources with exactly the same  distribution of treatment effect modifiers. 

Our proposed technique to assessing exchangeability of CATEs between studies applies the causal forest framework. To validly estimate CATEs, we make the following standard identifiability assumptions for causal estimands:
\begin{enumerate}
    \item \textit{Consistency,} where $Y_i = Y_i(1)Z_i + Y_i(0)(1-Z_i)$ for $i = 1, \dots, n_p + n_a$. This assumption indicates that the potential outcome under the observed treatment is the same as the observed outcome. 
    \item Mean \textit{unconfoundedness} within study $s$, $E[Y_i(Z) | Z_i = z, X_i = x, S_i = s] = E[Y_i(z) | X_i = x, S_i = s]$, for $z \in {0, 1}$. 
    \item \textit{Positivity} of treatment within study: $0 < Pr(Z_i = z | X_i = x, S_i = s) < 1$. 
\end{enumerate}

%\subsection{Causal Forest Framework}

%[Need an introduction to highlight we are choosing this causal forest framework over the other methods]

 %Recently, \cite{dandl2024makes} compared causal forests using this honest approach versus an adaptive approach (where the same sample is used for tree construction and effect estimation). They found that for smaller sample sizes and a scenario meant to imitate an RCT - adaptive forests performed better. However, we choose to continue to use Honest Causal Forests since we intend for our framework to be applied to either an RCT or observational study framework and do not limit our simulations to very small sample sizes. 

\subsection{Multi-Study Causal Forest (MCF)}

The Causal Forest framework was developed by \cite{wager2018estimation} for the purpose of treatment effect estimation using an adapted Random Forest algorithm. Just as a Random Forest \citep{breiman2001random} is an ensemble over many Classification and Regression Trees (CARTs), the Causal Forest framework is an ensemble over numerous Causal Trees \citep{athey2016recursive}.  An individual causal tree recursively partitions over the covariate space to determine the best split and corresponding criterion point by optimising for CATE heterogeneity. An `honest' approach was utilised to overcome the bias which may occur when the same data is used for the tree building process and for treatment effect estimation within the determined leaves. %The honest causal tree uses an in-sample expected mean square criterion (EMSE) to determine splits, 
%\begin{equation}
%-\hat{\mathbf{EMSE}}_{\tau}(S^{tr}, N^{est}, \prod) = \frac{1}{N^{tr}} \sum \hat{\tau}^2 (X_i; S^{tr}, \prod) - (\frac{1}{N^{tr}} + \frac{1}{N^{est}}) \sum_{l \in \prod} (V_{S ^{tr}}(l)). 
%\label{eq:emse}
%\end{equation}
%In Equation \ref{eq:emse}, $-\hat{\mathbf{EMSE}}_{\tau}$is the estimator of the EMSE of the treatment effect $\tau$, where $\tau$ is the treatment effect defined within a leaf node $l$, $\prod$ is a specific partition, $S^{tr}$ is the training sample used in the tree building process, $N^{tr}$ is the size of the training sample, $N^{est}$ is the size of the estimation sample and $V_{S ^{tr}}(l)$ is the within leaf-variance for the treated individuals.  The honest causal tree seeks a covariate split which corresponds to maximum heterogeneity between treatment effects of leaves and minimum variance of estimates within the individual leaves.
Using observed data within each leaf $(l)$ of a causal tree, the estimated CATE is computed as the difference in average outcomes between individuals in the treatment group and control group.

The Causal Forest framework was developed for a single study and is able to account for heterogeneity within the given study population. In this section, we describe how we adapt this single-study framework for borrowing across multiple data sources. When aggregating across multiple studies, several questions of interest need to be answered - 1) how does the pattern of treatment effect heterogeneity vary across the different studies? 2) how to assign higher weights to those observations from an auxiliary dataset deemed to be most similar to a target sample? This range of variation can range from perfectly homogeneous studies to presence of substantial heterogeneity between the studies. We introduce a data-driven framework to determine the extent of pooling between the different studies.

From an implementation perspective, we also exploit the Causal Forest framework's ability to weight the observations present in a given dataset. Propensity scores are utilised to balance the treatment and control groups when the covariate matrix $\mathbf{X}$ is observed. To evaluate the level of similarity between datasets, we also compute the Pearson Correlation Coefficient (PCC). In a dataset composed of both the primary and auxiliary datasets, we utilise a correlation coefficient and propensity scores for the purpose of introducing weights, separately as well as in combination. 
% Mention that we could use some other measure of concordance between the two sets of predicted values, provided they can be scaled to a weight on [0,1]
\subsubsection{Model}
%To begin with, let the primary data sample be denoted by $s_{pr}$. The training and test datasets from this primary dataset is denoted by $s_{tr}$ and $s_{te}$. 
%The auxiliary dataset we would like to pool with this original dataset is denoted by $s_{a}$ and a simple combination of the datasets is $s_{tr+a}$. 
A Classification Forest (for prediction of a binary outcome) is fit to predict the treatment assignment against covariates from the training dataset. Propensity scores can be used to weigh the auxiliary dataset and are computed against the set of baseline covariates from the dataset $s_{tr+a}$. Let the weights for the dataset $s_{tr}$ be equal to 1 and the weights for $s_{a}$ be  defined as 
\begin{equation}
    w_{i} = (Z_{i})\hat{\pi}_{tr}(\mathbf{X}_i) + (1- (Z_{i}))(1-\hat{\pi}_{tr}(\mathbf{X}_i)).
    \label{eq:prop}
\end{equation}
where $\hat{\pi}_{tr}$ is the propensity score fitted to the training dataset and evaluated at the covariates for individual $i$.
We define the Causal Forest framework fit to the training dataset be $CF_{tr}$, Causal Forest model fit to the auxiliary dataset be $CF_{a}$ and a model fit to a simple combination of the two datasets be $CF_{tr + a}$. We define the CATEs predicted from $CF_{tr}$ against the training dataset $s_{tr}$ to be $\tau_{CF_{tr}}(s_{tr})$. Similarly, the CATEs predicted from $CF_{tr+a}$ against the training dataset to be $\tau_{CF_{tr+a}}(s_{tr})$. We then determine the correlation coefficient $\rho$ between these estimated CATEs such that 
\begin{equation}
\rho_i = |\texttt{Corr}(\tau_{CF_{tr}}(s_{tr}), \tau_{CF_{tr + a}}(s_{tr})|.
\label{eq:corr}
\end{equation}
An alternative measure of concordance between two sets of predicted values may be used, provided that they can be scaled to a weight on [0, 1].
% Mention that we could use some other measure of concordance between the two sets of predicted values, provided they can be scaled to a weight on [0,1]
We define several Causal Forests to incorporate Equation \ref{eq:prop} and \ref{eq:corr} when working with dataset $s_{tr+a}$. $CF_{tr+a}$ is fit as a simple combination of datasets. With the use of Equation \ref{eq:prop} to weigh the auxiliary dataset, let the Causal Forest be denoted as $CF_{(tr+a)_{w_i}}$. Similarly, using Equation \ref{eq:corr} to weigh the auxiliary dataset, the Causal Forest is denoted as $CF_{(tr+a)_{\rho_i}}$. The Multi-study Causal Forest uses the product of Equations \ref{eq:prop} and \ref{eq:corr} to weigh the auxiliary dataset and is represented by $CF_{(tr+a)_{w_i *\rho_i}}$.  

%The above notation and definitions hold in the presence of a single auxiliary dataset, where the observations in $s_a$ are weighted against the observations in $s_{tr}$. All the observations in the primary data source are assigned weight $w_{tr} = 1$. In the presence of multiple supplementary datasets, weights are computed pairwise against the primary data source. To ensure that the relevance of the primary data source does not diminish (in comparison to the presence of multiple supplementary studies and larger volumes of datasets), we suggest a re-calibration of weights. 

%For the sake of ease in interpretation, we choose to utilise the case of a single auxiliary dataset (rather than multiple datasets) for evaluating the performance of the MCF in a simulation study and illustrating the method in an application.

\section{Simulation Study}

In this section, we evaluate the performance of our proposed method in its ability to determine when it's best to borrow from an auxiliary dataset against the target primary dataset. The target primary dataset and the auxiliary dataset each have a sample size of 500. Ten covariates $X_1, X_2, \dots, X_{10}$ were simulated as independent standard normals in both datasets, with $\mu = 0$ and $\Sigma = \Sigma_\rho$, where $\Sigma_\rho$ is a symmetric matrix with 1 on the diagonal and $\rho$ on every off-diagonal entry. We also generate heterogeneous treatment effects within the two datasets such that $\tau = \beta_{Y_Z}+ X\beta_{Y_{ZX}}$. A continuous outcome is generated within both datasets from a Normal distribution $Y \sim N(Y_{\mu}, 1)$, where $Y_{\mu} = X{\beta_{Y_X}} + \beta_{Y_Z}Z + \beta_{Y_{ZX}}ZX$.

In our simulation setup, we generate datasets in broadly three  categories to define varying degrees of heterogeneity between the two datasets 1) Primary = Auxiliary where the two datasets are simulated using the same parameters, 2) Primary $!=$ Auxiliary where the two datasets are simulated using different parameters and 3) Primary $\sim$ Auxiliary where the two datasets have partial similarity. We evaluate these scenarios 1) in the case of varied underlying propensity score models between the two datasets (where the treatment selection mechanism (i.e., propensity score model) differs between the primary and auxiliary dataset) and in the case of common underlying propensity score models between the primary and auxiliary dataset 2) in the presence of correlation among covariates in a dataset using $\Sigma_{\rho = 0.2}$ and absence of correlation among covariates in a datatset using $\Sigma_{\rho = 0}$. When the propensity score  models are different, the original dataset continues to use only $X_1$ but the auxiliary dataset uses $X_2^2, X_3$ and $X_4$. We also evaluated the case of common propensity score models, where both the original and the auxiliary dataset use only $X_1$ to generate propensity scores. we present these results in the appendix. A summary of relevant coefficient values for the category Primary != Auxiliary are presented in Table \ref{tab:coeffval_diff}. The coefficient values for the category when the Primary and Auxiliary datasets have the same parameters and similar parameters are presented in the Appendix (Tables \ref{tab:coeffval_same} and \ref{tab:coeffval_sim}).

\begin{table}[ht]
    \centering
    \begin{tabular}{ccccc}
    \hline
    & $\beta_{Y_{ZX}}$ & $\beta_{Z_X}$ & $\beta_{Y_Z}$ & $\beta_{Y_X}$  \\
     & Primary\\
    \hline
         Low & $(1, 0, \dots, 0)$  & 0.5 & 0.5 & $(1, 1, 1, 0, \dots, 0)$ \\
         Mid & $(1, 0, \dots, 0)$ & 0.5 & 0.5 & $(1, 1, 1, 0, \dots, 0)$ \\
         High & $(1, 0, \dots, 0)$ & 0.5 & 0.5 & $(1, 1, 1, 0, \dots, 0)$ \\
    \hline
     & Auxiliary\\
         Low &  $(0, 0.5, 0.5, 0.5, 0, \dots, 0)$ & 0.5 & 0.5 & $(1, 1, 1, 0, \dots, 0)$ \\
         Mid & $(0, 1.5, 1.5, 1.5, 0, \dots, 0)$ & 1.5 & 1.5 & $(1.5, 1.5, 1.5, 0, \dots, 0)$  \\
         High & $(0, 2, 2, 2, 0, \dots, 0)$ & 2 & 2 & $(2, 2, 2, 0, \dots, 0)$ \\    
    \end{tabular}
    \caption{Coefficient values used in the simulation study when the Primary and Auxiliary dataset use different parameters. This corresponds to a high level of between study heterogeneity}
    \label{tab:coeffval_diff}
\end{table}

We train our proposed method on a subset of the primary dataset and evaluate its performance against the test dataset (remaining 50\% of the primary dataset). Figure \ref{fig:diff_propscores} presents the performance results measured by the root mean squared error (RMSE) over 500 simulations. 

\subsection{Different propensity scores between studies}

\begin{figure}[h!]
        \centering
        \includegraphics[scale = 0.45]{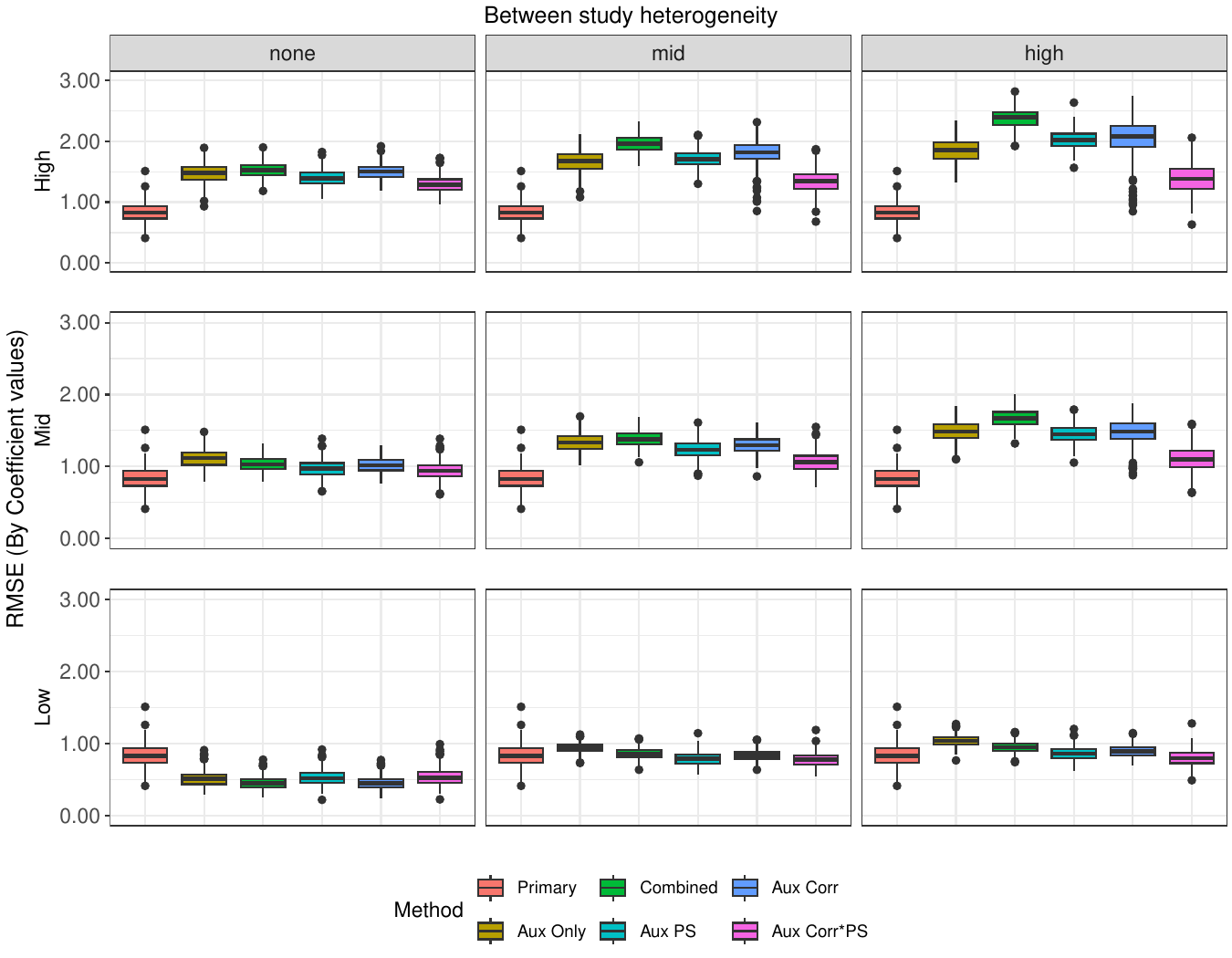}
        \caption{Simulation results are summarised to describe the performance of methods using the RMSE. The different columns correspond to varying levels of between study heterogeneity (from none to high) - the corresponding parameters are available in Tables \ref{tab:coeffval_diff}, \ref{tab:coeffval_same} and \ref{tab:coeffval_sim}. The different rows correspond to coefficient values also specified in the above Tables. The underlying data generating models specify correlation (0.2) among covariates within original and auxiliary datasets and also use different propensity score models}\label{fig:diff_propscores}
\end{figure}

Figure \ref{fig:diff_propscores} evaluates the performance of different methods when the propensity score models used in the data generating models are different and in the presence of correlation (at 0.2) among covariates within the the datasets. methods compared include Causal Forests built over the Primary dataset only (Primary), the Auxiliary dataset (Aux only), the Combined dataset (Combined) and weighted Causal Forest where the Auxiliary dataset is weighted using propensity scores (Aux PS), correlation (Aux Corr) and a combination (MCF).   %We start by summarising the results when there is no correlation among covariates within the datasets. 
When the Primary data is generated with the same parameters as the Auxiliary dataset (first column labelled as none - corresponding to no between study heterogeneity), the differences across a majority of the methods compared were minimal. This is especially true when the magnitude of coefficients are high and medium; in these two cases, using the Primary dataset alone corresponds to the lowest root mean squared error (RMSE). However, when the magnitude of coefficient is low, the RMSE is actually higher for a Causal Forest using only the Primary dataset. %This is true both in the presence and absence of correlation among covariates in the dataset. 
When the Primary dataset is generated with different parameters from the Auxiliary dataset (column labelled as high level of between study heterogeneity), combining the datasets results in significantly higher RMSE than simply using the Primary dataset. This is true at both high and medium coefficient values. The use of the MCF corresponds to a significantly lower RMSE than the RMSE for a combination of datasets (Combined). This difference is also present when only the propensity scores (Aux PS) or correlation (Aux Corr) is used to weigh the Auxiliary dataset. 

Similar results are observed in the presence of no correlation among covariates within the datasets and when the propensity score models are different between datasets (displayed in Figure \ref{fig:nocorrdiffps} in the Appendix). Figure \ref{fig:com_propscores} in the Appendix describes the performance of methods when common propensity scores are utilised in the data generating models for both datasets. Similar results are again observed in this set of plots comparing the MCF to the other methods. 

%\subsection{Scenario II: Equal propensity scores between studies}
%When the Original and Auxiliary datasets are different, the RMSE for Auxiliary only and Combined datasets are much higher than the RMSE for a Forest using only the Original dataset (nearly equivalent to the MCF). The range of RMSE values at a high magnitude level is much lower (max at a value of 5) when common propensity scores are utilised. The RMSE for the MCF is nearly equivalent to the original across magnitude levels for the coefficient. In the absence of correlation, the variation across the methods for the magnitude levels of the coefficient is quite limited. This is true regardless of whether original and auxiliary datasets are equivalent or distinct. 

\section{Breast Cancer Data Application}

We illustrate the use of the multi study causal forest in a realistic application using a randomised controlled trial (RCT) as the Primary data and an observational study as the Auxiliary data. These datasets are available within the R package \texttt{curatedBreastData} \citep{planey2015package} and were also utilised to demonstrate the application of the Multi-study R learner by \cite{shyr2023multi}. We follow a majority of the specifications of the datasets specified by \cite{shyr2023multi} (as described below) and then highlight the modifications we have made to the dataset. 

The purpose of this analysis is to characterise the heterogeneous treatment effects of two different neoadjuvant chemotherapy regimens for early breast cancer - anthracylene (A) versus taxane (T). The binary outcome of interest was pathological complete response - refers to the disappearance of all invasive cancer in the breast after completion of neoadjuvant chemotherapy ($Y = 1$) or otherwise ($Y = 0$). Two different studies ($K = 2$) were identified, where patients were administered the neoadjuvant chemotherapy. The first study, was a RCT of women aged between 18 and 79 years with stage II-III breast cancer \citep{martin2011genomic}. The second study, was an observational study of women with stage I-III breast cancer and were HER2 negative (HER2-) \citep{hatzis2011genomic}. For this analysis, five clinical characteristics (age, histology grade, HR+, PR+ and HER2+ status) and eight genes (SCUBE2, MMP11, BCL2, MYBL2, CCNB1, ACTB, TFRC, GSTM1) from Oncotype DX were selected to focus on. 

A challenge with illustrating treatment effect heterogeneity on real data is that we do not observe both potential outcomes, and so we simulated study-specific treatment effects. Using the potential outcome framework, the study-specific treatment effect is defined as $\tau_k = p_k^{(1)}(x) = p_k^{(0)}(x), k = \{1, 2\}$. These potential outcome probabilities were computed as linear functions of the five clinical features and eight genes on the logit scale. Also, note that HER2 status was not a confounder in study 2. In addition, to introduce between-study heterogeneity, random effects were assigned to the five clinical features and eight genes. Counterfactual outcomes $Y(a)$ were generated from a Bernoulli distribution with probability $p_k^{(a)}(x)$ and set $Y = Y(a)$.

For our analysis, the RCT dataset is referred to as the Primary dataset ($n_1 = 94$) and the observational study is the Auxiliary dataset ($n_2 = 168$). We then go on to randomly divide the Primary dataset to a training dataset ($n_{train} = 64$) and a test dataset ($n_{test} = 30$). In this analysis, we choose to use the Primary dataset as a target sample within the multi-study Causal Forest. The Multi-study Causal Forest was then utilised to determine the extent of auxiliary dataset to be used for the pooled dataset. The results of this method was compared to the Multi-study R Learner. As implemented by \cite{shyr2023multi}, the nuisance functions of the training data was estimated using Lasso with tuning parameters selected by cross validation. For the RCT, the propensity score was defined as 0.5 and estimates for the propensity score were obtained from the training data for the observational study. In order to estimate the membership probabilities in the multi-study R Learner (MSRL), a logistic regression model is fit to the training dataset's study labels. For both the multi-study models, we determine the heterogeneous treatment effect estimates and compute the MSE against the defined test dataset (a subset of the original RCT). 

\begin{figure}
    \centering
    \begin{subfigure}{0.9\textwidth}
    \centering
    \includegraphics[scale = 0.8]{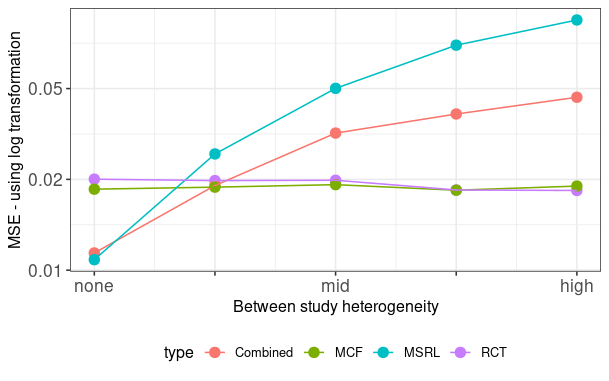}    
    \caption{Use of Training, Test and Auxiliary dataset}\label{fig:single}
    \end{subfigure}
    \begin{subfigure}{0.9\textwidth}
    \centering
    \includegraphics[scale = 0.8]{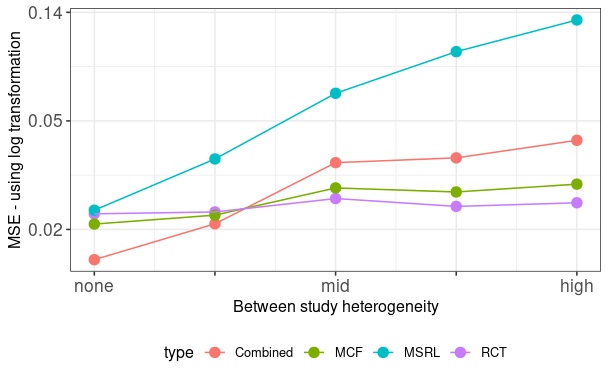}    
    \caption{Stacking of datasets to increase the underlying signal within the Primary dataset and encourage greater borrowing from the auxiliary dataset. This is done by repeating the Training and Test dataset but not the Auxiliary dataset.}\label{fig:double}
    \end{subfigure}
    \caption{Comparison of Multi-study Causal Forest (MCF) (including the combination of datasets, just the RCT and the MCF) against the Multi-study R learner (MSRL).}
    \label{fig:MSE_appl}
\end{figure}

\begin{figure}
    \centering
    \begin{subfigure}{\textwidth}
    \centering
    \includegraphics[scale = 0.95]{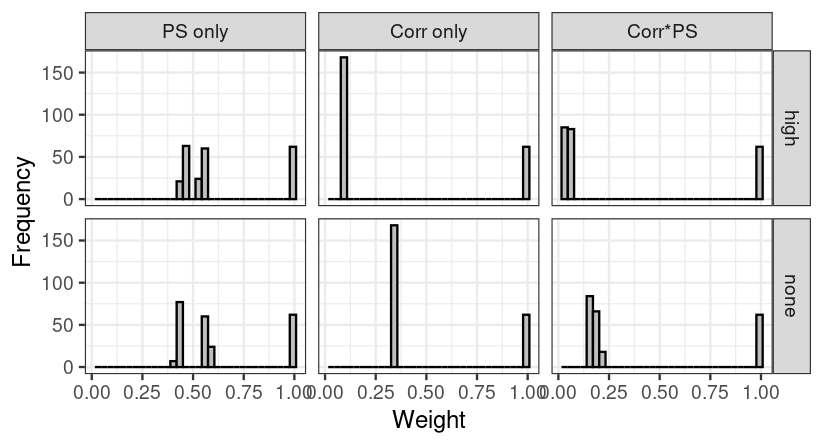}    
    \caption{Training, Test and Auxiliary dataset}\label{fig:sinwt}
    \end{subfigure}
    \begin{subfigure}{0.975\textwidth}
    \centering
    \includegraphics[scale = 0.95]{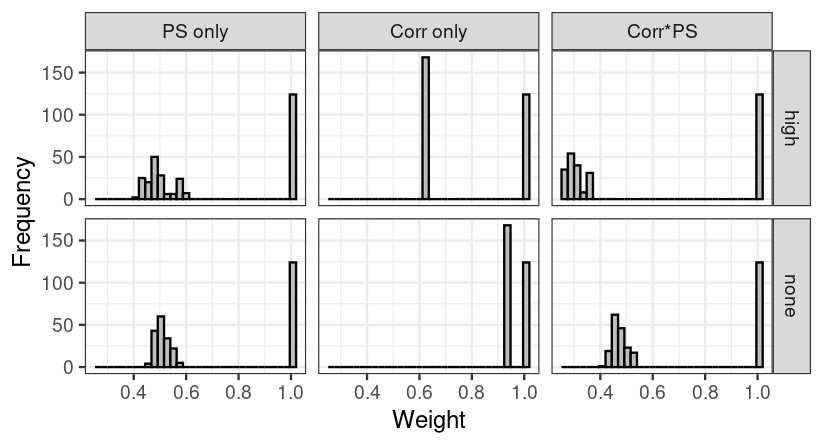}    
    \caption{Repeated training and test dataset but not the Auxiliary dataset.Stacking of datasets to increase the underlying signal within the Primary dataset and encourage greater borrowing from the auxiliary dataset. This is done by repeating the Training and Test dataset but not the Auxiliary dataset.}\label{fig:dowt}
    \end{subfigure}
    \caption{Distribution of weights within the Multi-study Causal Forest}
    \label{fig:weights_appl}
\end{figure}

Figure \ref{fig:MSE_appl} compares the performance of the approaches against the mean-squared error (MSE). In Figure \ref{fig:single}, when there is no between study heterogeneity, the multi-study R Learner and a combination of datasets for use in the Causal Forest outperforms the multi-study Causal Forest and when only the RCT is utilised. As the between-study heterogeneity increases, the MCF rightly identifies that pooling the datasets is not the right choice here. This is unlike the MSRL which continues to pool regardless of the disadvantages this poses. In this scenario, the MCF uses a rather small dataset and the underlying weights are described in Figure \ref{fig:sinwt}. Across the range of between study heterogeneity values, the computed correlation is rather low and along with the propensity scores, the weights for pooling the auxiliary dataset are lower than 0.25. Accordingly, the MCF utilises only a very small subset of the greater auxiliary dataset along with the training dataset. In order to increase the underlying signal, the rows of both the training and test datasets are duplicated and stacked to create larger datasets. With this greater underlying signal, the correlation between the datasets increases significantly in the scenario when there is no between study heterogeneity. However, after accounting for the propensity score, the weights for the MCF are only between 0.4 to 0.6, which indicates only around a half of the auxiliary dataset is pooled for use by the MCF. This proportion is accurately far lower for "high" between study heterogeneity. Also, in this scenario, with stacked datasets, the MSE for MSRL continues to increase as between-study heterogeneity increases. 

\section{Discussion}

This article proposes the Multi-Study Causal Forest framework for borrowing from supplementary data in the presence of treatment effect heterogeneity. This flexible approach is able to identify when differences in CATE estimates (between studies) may be entirely attributed to differences in treatment effect-modifiers (when covariate-adjusted exchangeability holds) and when other reasons are driving the differences. This helps to determine the extent to which data should be pooled from an auxiliary data source. The simulation study demonstrates the ability of the method to correctly choose to borrow when exchangeability holds by utilising the higher assigned weights to the auxiliary dataset. Lower weights correspond to a decision of the framework to recommend a strategy of limited (or no) borrowing. In general, this method is scalable given that its individual components are scalable (including the Causal Forest framework, the use of a Random Forest framework to compute propensity score weights and calculation of the RMSE). Given our choice of utilising a Causal Forest framework, the MCF inherits the advantages (e.g., ability to accommodate non-linear relationships) and limitations of the method (e.g., does not work for a very small sample size due to the need for minimum sample sizes to be set in the tree splitting process). However, other data-driven approaches which focus on estimating CATEs can also be utilised, if they offer the ability to pass weights for observations within the study population.

We make several assumptions in the development of the MCF framework including 1) this methodology assumes that researchers have access to individual-level data, measuring a common set of covariates, from multiple sources, 2) the same two treatment arms in the supplementary studies as in the primary study 3) a common set of covariates under consideration in different data sources. We did not discuss any scenarios where the data sources contain different number of covariates or in the presence of missingness. There may also be cases when data is only available on a single arm in the auxiliary study. We also work within only one single-study framework, the non-parametric Causal Forest framework, for estimating CATEs but we can potentially utilise other data-driven frameworks for evaluating treatment effect heterogeneity. A related line of work is federated learning across multiple data sources, which may be needed when it is not feasible to share individual-level data. However, it may be possible to leverage models and parameters from other sites to incorporate it in the existing framework. Additionally, we only seek to borrow from a single supplementary data source and do not offer an approach to integrate data from multiple (greater than one) sources. We leave such extensions to future work.

\section*{Acknowledgements}

AV is an employee of GlaxoSmithKline (GSK). The views in this manuscript are solely that of the authors and do not represent GSK in any way. JW reports no conflicts of interest.

\bibliography{refs}
\bibliographystyle{plainnat}

\section*{Appendix}

\begin{table}[ht]
    \centering
    \begin{tabular}{ccccc}
    \hline
    & $\beta_{Y_{ZX}}$ & $\beta_{Z_X}$ & $\beta_{Y_Z}$ & $\beta_{Y_X}$  \\
     & Primary\\
    \hline
         Low & $(1, 0, \dots, 0)$  & 0.5 & 0.5 & $(1, 1, 1, 0, \dots, 0)$  \\
         Mid & $(1, 0, \dots, 0)$ & 0.5 & 0.5 & $(1, 1, 1, 0, \dots, 0)$ \\
         High & $(1, 0, \dots, 0)$  & 0.5 & 0.5 & $(1, 1, 1, 0, \dots, 0)$\\
    \hline
     & Auxiliary\\
         Low &  $(1, 0, \dots, 0)$ & 0.5 & 0.5 & $(1, 1, 1, 0, \dots, 0)$\\
         Mid & $(1, 0, \dots, 0)$ & 1.5 & 1.5 &  $(1.5, 1.5, 1.5, 0, \dots, 0)$ \\
         High & $(1, 0, \dots, 0)$  & 2 & 2 & $(2, 2, 2, 0, \dots, 0)$ \\    
    \end{tabular}
    \caption{Coefficient values used in the simulation study when the Primary and Auxiliary dataset use the same parameters. This corresponds to no between study heterogeneity.}
    \label{tab:coeffval_same}
\end{table}

\begin{table}[ht]
    \centering
    \begin{tabular}{ccccc}
    \hline
    & $\beta_{Y_{ZX}}$ & $\beta_{Z_X}$ & $\beta_{Y_Z}$ & $\beta_{Y_X}$  \\
     & Primary\\
    \hline
         Low & $(1, 0, \dots 0)$  & 0.5 & 0.5 & $(1, 1, 1, 0, \dots, 0)$  \\
         Mid & $(1, 0, \dots 0)$ & 0.5 & 0.5 & $(1, 1, 1, 0, \dots, 0)$ \\
         High & $(1, 0, \dots 0)$ & 0.5 & 0.5 & $(1, 1, 1, 0, \dots, 0)$\\
    \hline
     & Auxiliary\\
         Low & $(0.25, 0.25, 0.25, 0.25, \dots )$  & 0.5 & 0.5 & $(1, 1, 1, 0, \dots, 0)$\\
         Mid & $(0.375, 0.375, 0.375, 0.375, \dots )$ & 1.5 & 1.5 & $(1.5, 1.5, 1.5, 0, \dots, 0)$ \\
         High & $(0.5, 0.5, 0.5, 0.5, \dots)$  & 2 & 2 & $(2, 2, 2, 0, \dots, 0)$\\    
    \end{tabular}
    \caption{Coefficient values used in the simulation study when the Primary and Auxiliary dataset use similar but not the same parameters. This corresponds to a medium level of between study heterogeneity.}
    \label{tab:coeffval_sim}
\end{table}

\begin{figure}[h!]
        \centering
        \includegraphics[scale = 0.45]{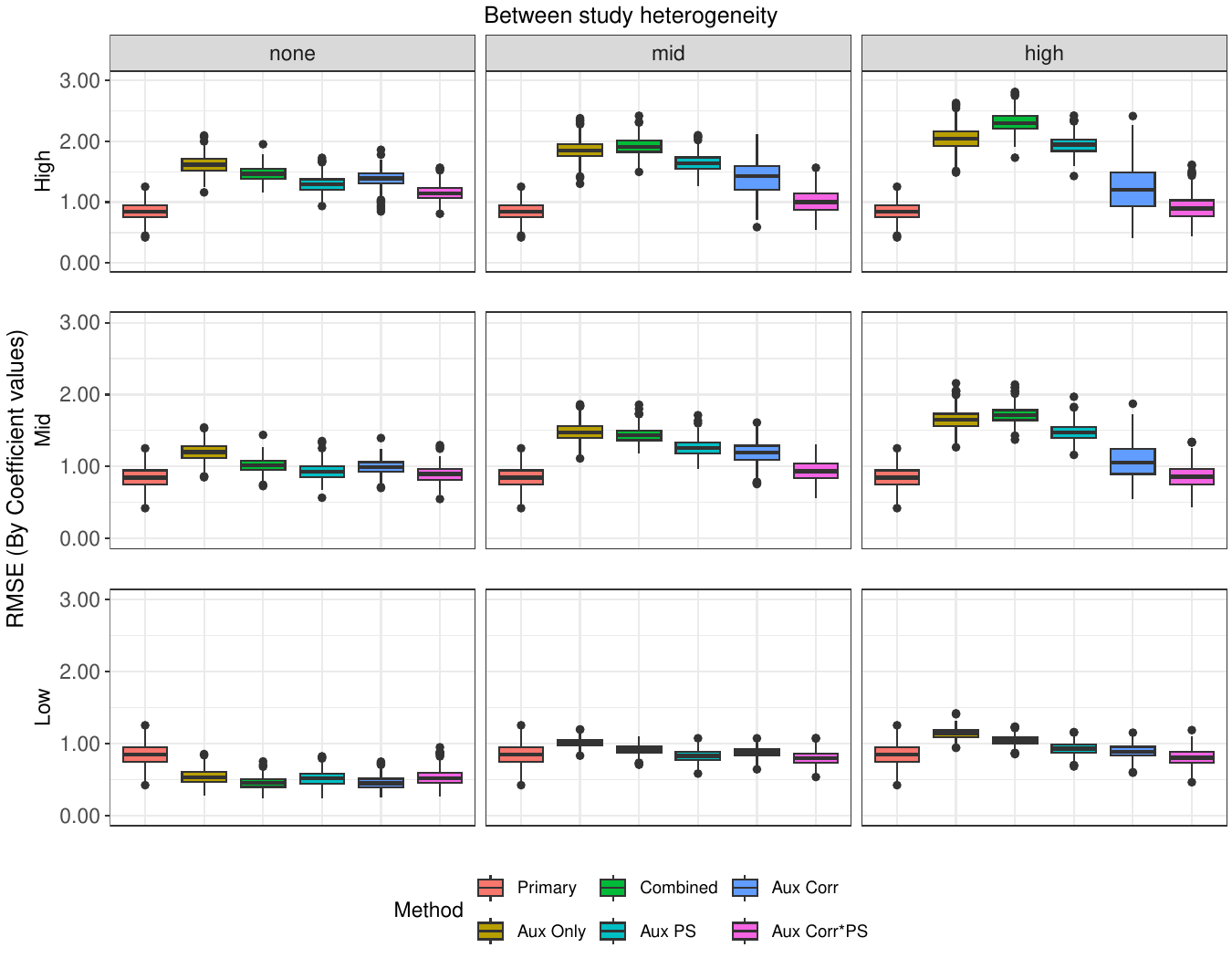}
        \caption{Simulation results summarised to describe the performance of methods using the root mean squared error (RMSE). The different columns correspond to varying levels of between study heterogeneity (from none to high) - the corresponding parameters are available in Tables \ref{tab:coeffval_diff}, \ref{tab:coeffval_same} and \ref{tab:coeffval_sim}. The different rows correspond to coefficient values also specified in the above Tables. The underlying data generating models specify no correlation among covariates within original and auxiliary datasets and also use different propensity score models}\label{fig:nocorrdiffps}
    \end{figure}

\begin{figure}
    \centering
    \begin{subfigure}[t]{0.95\textwidth}
        \centering
        \includegraphics[scale = 0.475]{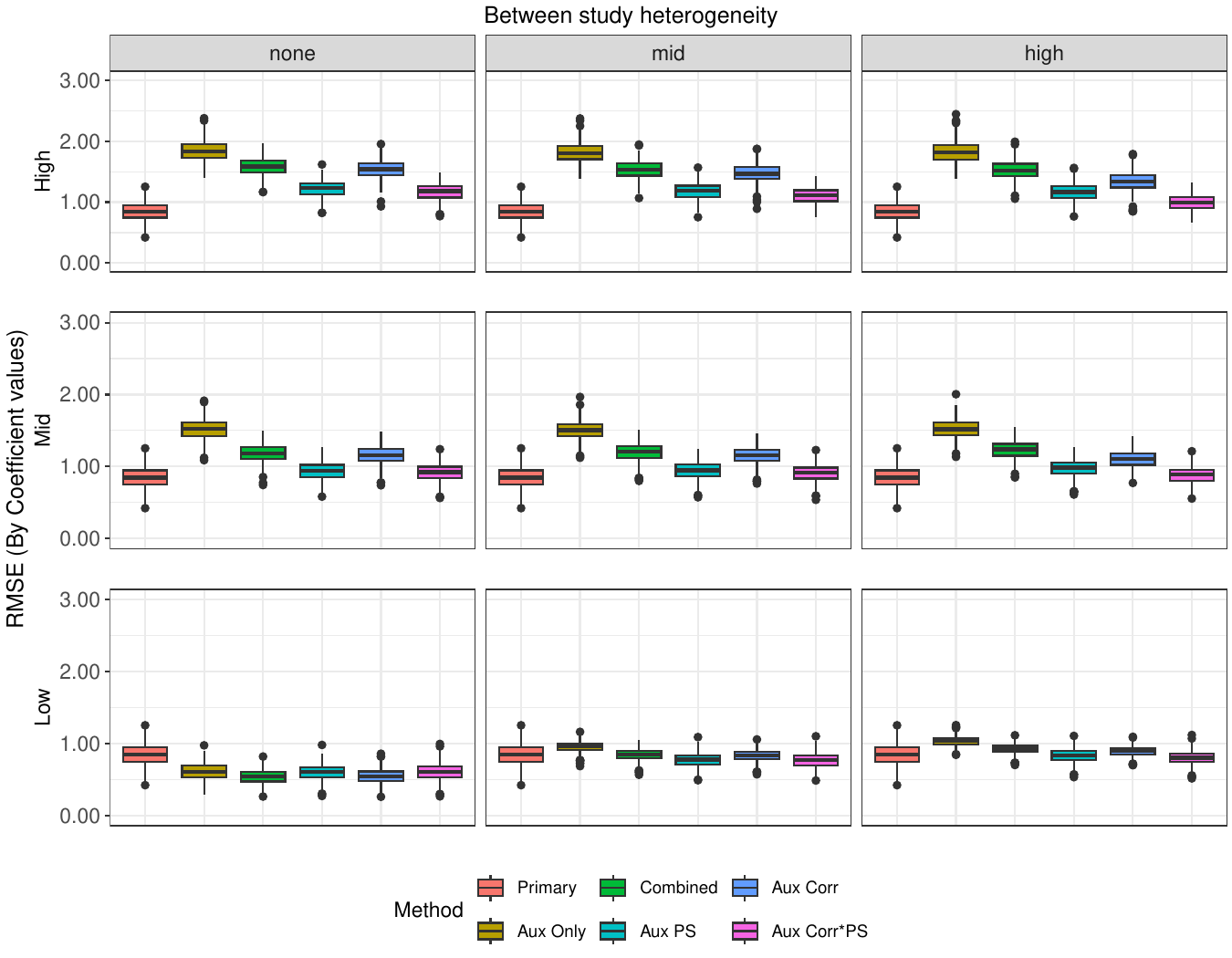}
        \caption{No correlation among covariates within Primary and Auxiliary datasets}
    \end{subfigure}%
    
    \begin{subfigure}[t]{0.95\textwidth}
        \centering
        \includegraphics[scale = 0.475]{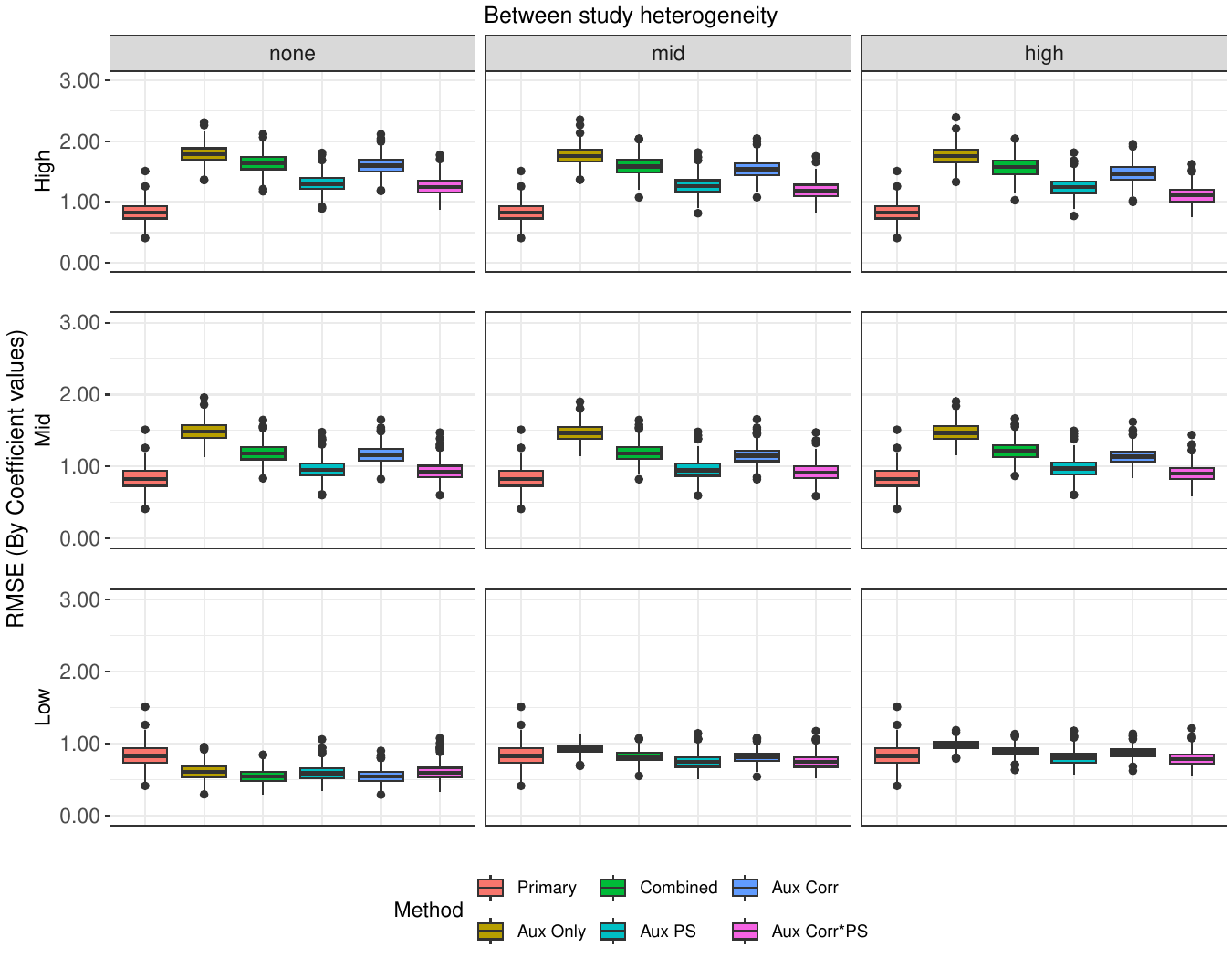}
        \caption{Presence of correlation (0.2) among covariates within both the Primary and Auxiliary datasets}
    \end{subfigure}
    \caption{Simulation results summarised to describe the performance of methods using the RMSE. The different columns correspond to varying levels of between study heterogeneity (from none to high) - the corresponding parameters are available in Tables \ref{tab:coeffval_diff}, \ref{tab:coeffval_same} and \ref{tab:coeffval_sim}. The different rows correspond to coefficient values also specified in the above Tables. The underlying data generating models use common propensity score models.}\label{fig:com_propscores}
\end{figure}

\end{document}